\documentclass[intlimits,twoside,a4paper]{article}
\usepackage{amsmath,amssymb}
\usepackage{graphicx}
\usepackage{wrapfig}
\usepackage[T2A]{fontenc}
\usepackage[cp1251]{inputenc}
\usepackage{siunitx,color}

\usepackage{cmpj2}

\issue{2013}{16}{1}{13801}

\doinumber{10.5488/CMP.16.13801}

\title[Structural properties of Zr--Nb alloys]%
{Ab-initio calculations for structural properties \\ of Zr--Nb alloys}

\author[V.O. Kharchenko, D.O. Kharchenko]{V.O. Kharchenko\refaddr{label1,label2},
		D.O. Kharchenko\refaddr{label1}}
\addresses{
\addr{label1} Institute of Applied Physics of the National Academy of Sciences of Ukraine, \\ 58~Petropavlivska~St., 40030~Sumy, Ukraine
\addr{label2} Institute of Physics, University of Augsburg, 1~Universit\"{a}tsstra{\ss}e, 86135 Augsburg, Germany
}

\date{Received  August 27, 2012, in final form November 9, 2012}
\authorcopyright{V.O. Kharchenko, D.O. Kharchenko, 2013}

\newcommand{\dc}{\si{\degreeCelsius}}

\begin{document}

\maketitle

\begin{abstract}

Ab-initio calculations for the structural properties of Zr--Nb alloys at
different values of the niobium concentration are done at zero temperature.
Different cases for Zr--Nb alloys with unit cells having BCC and HCP
structures are considered. Optimal values of the lattice constants are
obtained. Critical value for the niobium concentration corresponding to the
structural transformation HCP~$\rightarrow$~BCC at zero temperature is determined.
Electronic densities of states for two different structures with niobium
concentrations $12.5\%$ and $25\%$ having HCP and BCC structures, accordingly,
are studied.


\pacs 61.50.Lt, 61.72.Bb, 61.72.jj, 61.72.S-, 71.15.Dx, 71.15.Nc
\end{abstract}

\section{Introduction}

Zirconium alloys are known to be very promising materials for atomic industry
and power engineering starting from the recent six decades and are of  very special
practical importance. Zirconium and zirconium-niobium alloys are
widely used as constructions materials in nuclear reactors~\cite{MPH1,MPH2,MPH3}. The role of construction materials lies in a stability
support for the whole operating period of geometry of active zones, fuel
assemblies, heat-generating elements, etc. Therefore, the key problem for
material science security of modern nuclear power and for the future power
engineering is the study of micro-structure evolution and its effect on the
behavior of physical and mechanical characteristics of material. The materials used
in nuclear power engineering contain solute species in more or less dilute
proportions (1$\%$ in pressure vessel steels and in cladding materials, Zr
alloys). It is known that zirconium has two configurations: $\alpha$-phase
characterized by the HCP lattice that is stable at a low enough temperature
($<{863}\dc$), and $\beta$-phase with BCC lattice at high temperature. Niobium
has BCC lattice. Zirconium-niobium alloys characterized by $\alpha$-phase are
used as construction material of the active zone of nuclear reactors due to
a beneficial mix of nuclear, mechanical and corrosive properties
\cite{MPH1,MPH2,MPH3}.

Alloys Zr--$x$Nb with a high percentage of niobium ($x > 5 \%$) have a high
capacity for hydrogenation which causes deterioration of corrosion properties.
That is why zirconium-niobium alloys with small
niobium concentration ($< 5 \%$) can be advantageously used in the reactor metallurgy. In the nuclear reactor,
Hydrogen diffuses inside an alloy due to corrosion processes. When the Hydrogen
concentration exceeds the solubility limit, it stands out as a fragile
hydride phase that limits the usage of an alloy in the reactor. Hence, the
study of Zr--Nb alloys having small and high niobium concentrations, together
with processes of hydride formation in zirconium-niobium alloys is an urgent
problem in the novel material science. Investigations of the hydride phase formation
processes in alloy Zr--Nb that contains $20\%$ and $40\%$ of niobium were made in
\cite{MPH3,MPH4}. To study such a process in alloys having high niobium
concentration, the plane strain theory was used in reference~\cite{MPH6}. In
reference~\cite{MPH7} diffusion coefficients of Zr and Nb atoms in BCC alloys
having niobium concentrations $5.5\%$, $16.3\%$ and $28.1\%$ were studied. It was found
that zirconium and niobium atoms diffuse in an alloy according to the mono-vacancy
mechanism. The latter is defined by the specification of lattice dynamics.
Diffusion of hafnium and niobium atoms in coarse-grained BCC alloy Zr--Nb with
$19\%$ niobium was studied in reference~\cite{MPH8}. It was determined that diffusion
coefficient of niobium is less than the hafnium one. Corrosion behavior of
Zr--Nb alloy with niobium concentration variation from $0.02\%$ to $20\%$ was considered
in~\cite{MPH9}. It was shown that corrosion rate becomes smaller with
a decrease in both bulk of the $\beta$-phase and niobium concentration in
$\alpha$-phase. The authors have determined that the most important factor that reduces the
corrosion rate is the precipitate of the $\beta$-Nb particles and the
corresponding decrease of niobium in the zirconium matrix. Experimental
investigations of $\beta$-phase formation with niobium concentration
of $20\div88\%$ at irradiation were made in
references~\cite{MPH10,MPH11,MPH12,MPH14}. Experimental results for the
structural properties $\alpha$- and $\beta$-phases in Zr--Nb alloys with
variation in niobium concentration at high temperature ($\sim700$~K) are
presented in references~\cite{fig1Ref2,JAC2,JAC3}. Structural properties of
Zr--$12.5\%$Nb and Zr--$25\%$Nb were theoretically considered in~\cite{MPH,vesnik}. However, the question regarding the critical value of
niobium concentration at the transition from $\alpha$-phase toward
$\beta$-phase (structural transformation HCP~$\rightarrow$~BCC) in
zirconium-niobium alloys still remains open.

In this work, we aim to study structural properties of zirconium-niobium
alloys with different niobium concentrations from the first principles at low
(zero) temperature. Model structures of Zr--Nb alloys with BCC and HCP
lattices will be considered. The main goal of this paper is to define optimal
lattice constants of Zr--Nb alloys with BCC and HCP lattices for different
niobium concentrations in the alloy. We shall define the critical value of
niobium concentration that corresponds to the structural transformation
HCP~$\rightarrow$~BCC. A quantitative comparison with the known experimental data
will be done.

The work is organized in the following manner. In section~2, we present the models
for the structures of Zr--Nb alloys and describe the research methods. In
section~3, we discuss the structural properties and obtain optimal lattice constants
for the structures studied. The analysis of the total energy and electronic density of
states is done in section~4. We conclude in the last section.

\section{Models of the structures studied and research methods}

It is known that zirconium at low temperatures ($< 863\dc$) is characterized
by $\beta$-phase (HCP structure), whereas niobium has BCC structure. Hence, it
is only natural to expect that alloys Zr--$x$Nb will have HCP structure at low
niobium concentration $x$ whereas BCC structure will be stable for large $x$.
Since zirconium and niobium are different only for one electron and are
characterized by close package structures, one can expect that niobium atoms in
the HCP zirconium lattice or zirconium atoms in the BCC niobium lattice are
substitutional atoms. In reference~\cite{MPH18} the authors have studied the super-cell of 48
atoms of HCP zirconium which has a vacancy and a niobium atom in interstitial
position placed near the vacancy. It was shown that during relaxation
processes, niobium atom takes a vacancy position. The latter confirms that the atoms
of Zr or Nb in Zr--Nb alloys are substitutional ones.

In this paper we consider the Zr--$x$Nb alloys with BCC and HCP lattices
where the niobium concentration $x$ takes the values $x=3.125\%$, $6.25\%$,
$12.5\%$, $25.0\%$, $50.0\%$ and $x=75.0\%$ for both types of lattices, and
additionally $x=87.5\%$, $93.75\%$ and $x=96.875\%$ for BCC alloys. In such a
case, structures with niobium concentrations $3.125\%$ and $96.875\%$ are
characterized by the largest number of atoms in the unit cell, 32. Hence, for
the above mentioned values of the niobium concentration $x$ in the alloy Zr--$x$Nb with
the unit cell of 32 atoms, a number of niobium atoms is: 1, 2, 4, 8, 16, 24,
28, 30 and 31, respectively.

Model structures of BCC alloys with 2, 16 and 30 niobium atoms in the unit cell
of 32 atoms are characterized by the 221 symmetry space group; other
structures have the 123 group of symmetry. All the studied structures of
Zr--Nb alloys with HCP lattice are characterized by the 187 space group. The
number of nonequivalent atoms of zirconium and niobium that gives a number of
symmetry operations for each model of BCC and HCP structure is shown in table~\ref{table1}.

\begin{table}[!ht]
\caption{The number of niobium atoms in the unit cell of 32 atoms (\# Nb$_\mathrm{UC}$)
and the number of nonequivalent atoms of zirconium (\# NEA$_\mathrm{Zr}$) and niobium
(\# NEA$_\textrm{Nb}$) for each model structure with BCC and HCP lattices.\label{table1}}
\vspace{2ex}
\begin{center}
\begin{tabular}{c|c|c|c}
  \hline \hline
  \# Nb$_\mathrm{UC}$ & Type of &  \# NEA$_\mathrm{Zr}$ & \# NEA$_\textrm{Nb} $\\
	& the lattice &    &  \\
  \hline\hline
  1 & BCC & 8 & 1\\
	& HCP & 9 & 1\\
  \hline
  2 & BCC & 4 & 1\\
	& HCP & 5 & 1\\
  \hline
  4 & BCC & 3 & 1\\
	& HCP & 3 & 1\\
  \hline
  8 & BCC & 2 & 1\\
	& HCP & 2 & 2\\
  \hline
 16 & BCC & 1 & 1\\
	& HCP & 2 & 2\\
  \hline
 24 & BCC & 1 & 2\\
	& HCP & 2 & 2\\
  \hline
 28 & BCC & 1 & 3\\
	& HCP & 1 & 3\\
  \hline
 30 & BCC & 1 & 4\\
  \hline
 31 & BCC & 1 & 8\\
  \hline \hline
\end{tabular}
\end{center}
\end{table}

All calculations of structural and electronic properties for the model alloys,
shown in table~\ref{table1}, were made within the framework of density functional
theory (DFT)~\cite{16epjb} using the linearized augmented plane wave (LAPW)
method, which is implemented in software packages Wien2k~\cite{17epjb,17epjb_}. This
method self-consistently includes basic and valence electrons and is widely
used to  calculate the band structure of solids~\cite{17epjb,17epjb_}. For
all calculations, the Muffin-tin radii ($R_\mathrm{MT}$) of both Zr and Nb were
taken to be $2.1$ atomic units. The basis function was
expanded up to $R_\mathrm{MT}\times k_\mathrm{max}=7$, where $R_\mathrm{MT}$ is the smallest radius of
the muffin-tin (MT) spheres and $k_\mathrm{max}$ is the maximal value of the
reciprocal lattice vectors. Wave function expansion inside the atomic spheres
was done up to $l_\mathrm{max}=10$. Such values for $k_\mathrm{max}$ and $l_\mathrm{max}$ are
standard. Deviations from these values should be considered in calculating the
systems with elements having large difference in masses. Integration over the
Brillouin zone was performed using 1000 $k$-points, which is enough for the
calculation of metallic structures. The criterion for stopping the iteration process
was the converge for the total energy and the charge of the crystal to less than
$0.0001$~Ryd and $0.001\, e^{-}$, respectively. All calculations were made using the generalized gradient approximation (GGA) with parametrization (PBE)
\cite{19epjb}.

\section{Structural relations}

In this section the main results are shown for structural properties of the
structures studied. The optimization procedure was used to define the optimal values of the lattice constants for
each structure of Zr--Nb alloy with BCC and HCP crystals.

\subsection{BCC lattice}

\begin{wrapfigure}{i}{0.5\textwidth}
\centering
\includegraphics[width=0.48\textwidth]{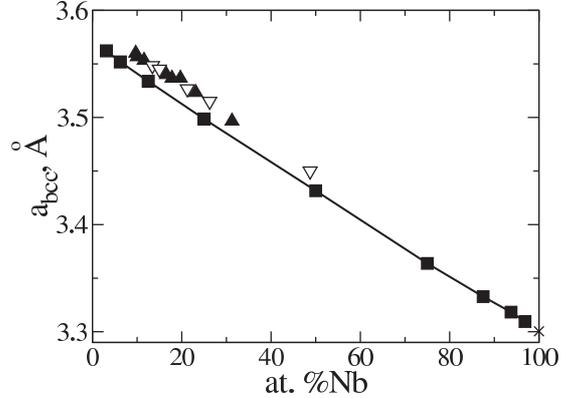}
\caption{The dependence of the lattice parameter $a_\mathrm{bcc}$ for the BCC
Zr--Nb alloys versus niobium concentration.} \label{fig1}
\end{wrapfigure}

For Zr--$x$Nb alloys with BCC lattice, the optimization procedure was made in the
standard manner. For each value of niobium concentration $x$, we have defined
the total energy for the unit cell of the crystal with varying in the unit cell
volume. The obtained data were approximated by the equation of state
\cite{Murnaghan}. The minimum on the dependence of the total energy of unit
cell versus unit cell volume corresponds to the optimal value of the unit cell
volume. Following the standard definition for the volume of the unit cell of
the BCC-crystals $V_\mathrm{bcc}=a_\mathrm{bcc}^3$, the optimal value for the lattice constant
$a_\mathrm{bcc}$ was determined. The dependence of the optimal lattice constant $a_\mathrm{bcc}$
on the niobium concentration $x$ for BCC alloys Zr--$x$Nb is shown in figure~\ref{fig1}.

Here, the obtained theoretical results are shown with the help of filled squares.
Empty and filled triangles correspond to the experimental data (see references~\cite{fig1Ref1} and \cite{fig1Ref2}, respectively). The known value for the
lattice constant of pure BCC niobium is shown by symbol $\times$. In
figure~\ref{fig1} it is seen that with an increase in the niobium concentration in
the BCC Zr--$x$Nb alloy the lattice constant decreases according to the linear
law. This result is consistent with the well known approximate empirical rule,
called in metallurgy Vegard's law~\cite{vegard,vegard_2}. According to this rule, at
constant temperature, a linear dependence is realized between the crystal lattice parameter
of an alloy and the concentrations of the constituent elements.
Thus,
%
for the studied BCC Nb$_x$Zr$_{1-x}$ alloy, where zirconium atoms are
substitutional ones in the BCC niobium lattice, the relation between lattice
parameters for the pure zirconium and niobium and its alloy is given as
follows: $a_\textrm{ZrNb}^\mathrm{bcc}=x\cdot a_\textrm{Nb}^\mathrm{bcc}+(1-x)\cdot a_\textrm{Zr}^\mathrm{bcc}$. Here,
the value for the lattice constant of pure BCC niobium is known,
$a_\textrm{Nb}^\mathrm{bcc}=3.3004$~{\AA}, whereas $a_\textrm{Zr}^\mathrm{bcc}=3.5634$~{\AA}~ is a fitting
parameter (at low temperatures zirconium is characterized by the HCP lattice).

It should be noted that in the interval of niobium concentration $x$ from $10\%$
to $50\%$ in the BCC Zr--$x$Nb alloy, the experimental data slightly exceed the
obtained theoretical data. This is only natural, because theoretical calculations
were performed at low (zero) temperature, while the presented experimental
results were obtained at evaluated temperatures~\cite{fig1Ref1,fig1Ref2}. It was
experimentally shown that in the Zr--$x$Nb crystal at evaluated temperatures
($\sim700$~K), BCC phase is realized at evaluated and large values for the
niobium concentration, i.e., at.$\%$Nb $>10\%$~\cite{fig1Ref1,fig1Ref2}. In
figure~\ref{fig1} we also show the values for a lattice constant of BCC alloy at small
niobium concentrations. In the next section we analyse the total
energy values of the crystals at low (zero) temperature and define the
minimal value of the niobium concentration in the alloy Zr--$x$Nb, where the crystal
is characterized by BCC lattice.

\subsection{HCP lattice}

\begin{figure}[!t]
\centering
\includegraphics[width=0.45\textwidth]{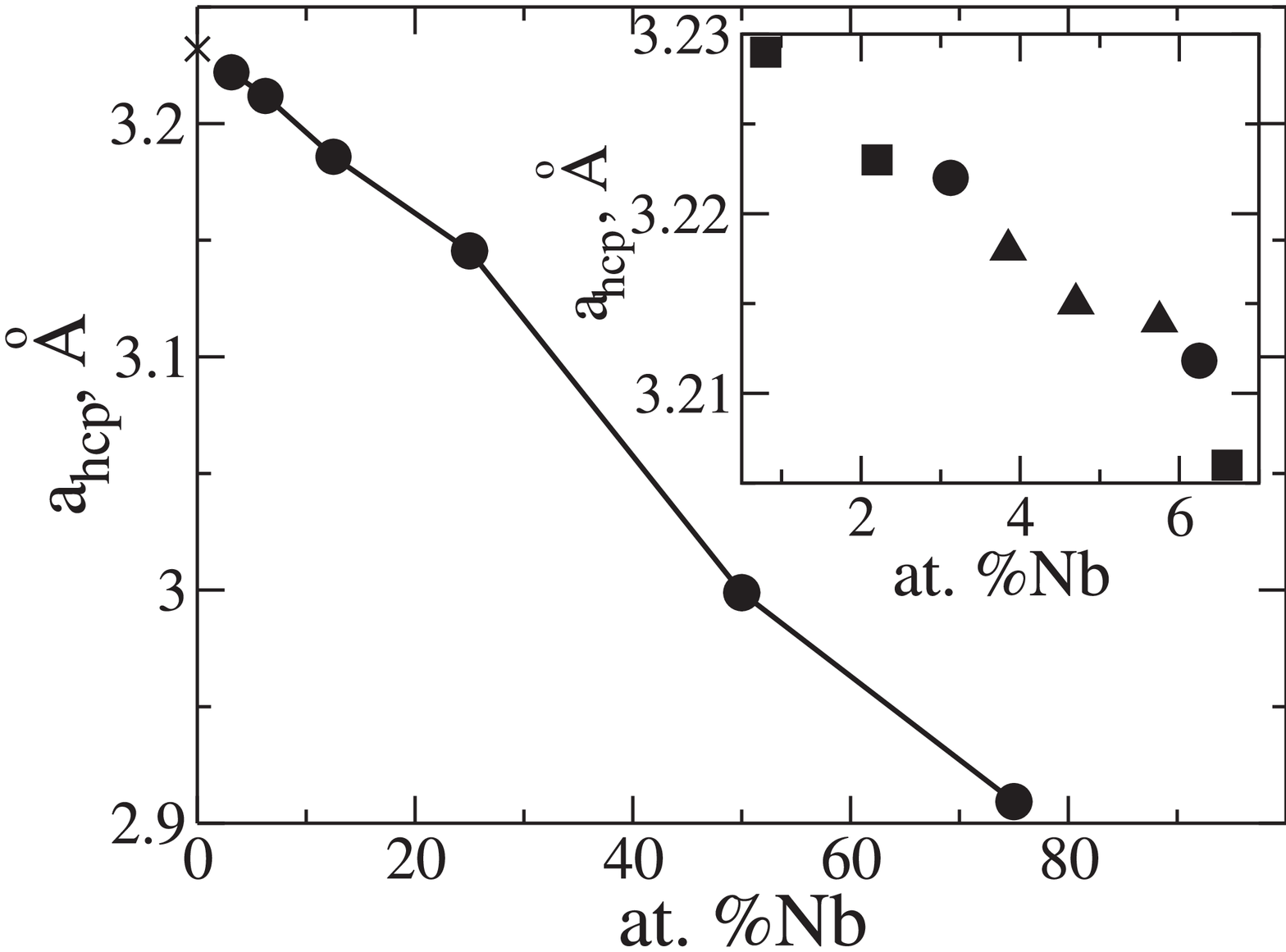}
\hspace{0.05\textwidth}
\includegraphics[width=0.45\textwidth]{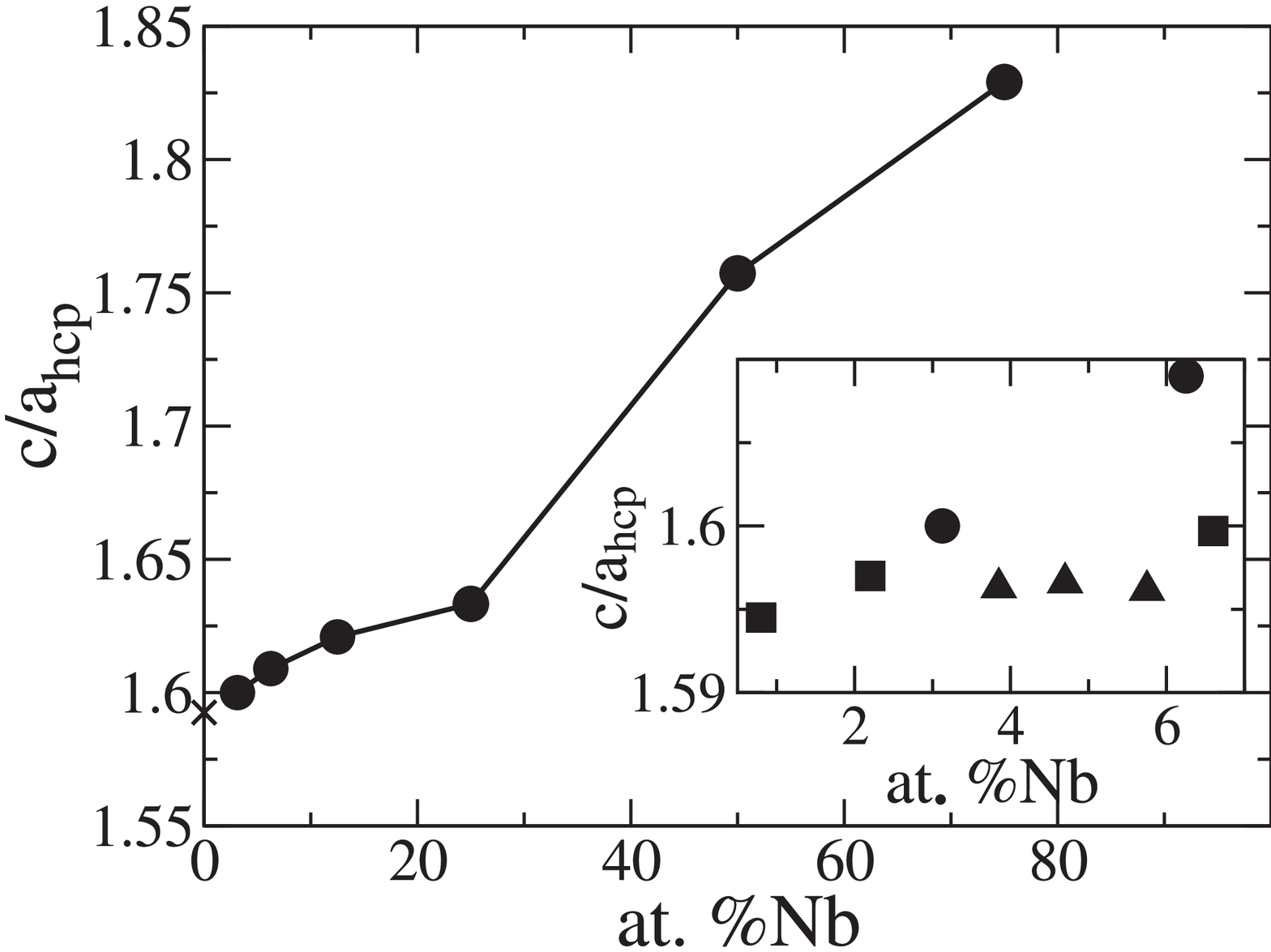}\\
\hspace{0.05\textwidth}(a)\hspace{0.5\textwidth} (b)
\caption{Dependencies of the lattice
parameter $a_\mathrm{hcp}$ (a) and structural relation $c/a$ (b) for the HCP
Zr--Nb alloys versus niobium concentration.} \label{fig2}
\end{figure}

Now, let us consider Zr--$x$Nb alloys with HCP lattice. The optimization
procedure allowing us to find optimal lattice constants $a_\mathrm{hcp}$ and $c_\mathrm{hcp}$
for such crystals is more difficult than in the previous case for BCC lattice.
This is because for the HCP alloy, one should define a global minimum of the
total energy of a crystal as a function of both the volume of the unit cell and the
structural relation $c/a$. To this end, we have proceeded in the following
manner. For each value of niobium concentration, a test value of structural
relation $c/a$ was fixed. The standard optimization procedure, like in the case
of BCC lattice, was used to define the optimal volume of the unit cell corresponding
to the minimum of the total energy as a function of the unit cell volume,
$E_\mathrm{min}^{(i)}\big((c/a)_i\big)$, where $i=1\ldots N$, $N$ is the  number of test values
for the structural relation $c/a$. The obtained empiric data as the dependence
$E_\mathrm{min}(c/a)$, were approximated by the functional dependence. The minimum of this
dependence corresponds to the optimal value of the structural relation
$(c/a)_\mathrm{opt}$ for each value of niobium concentration $x$ in the HCP
 Zr--$x$Nb alloy. Next, for the determined $(c/a)_\mathrm{opt}^{\%\textrm{Nb}}$, the standard
optimization procedure was made to determine a global minimum of the total energy
and the optimal unit cell volume, respectively. Using the standard definition
for the volume of the HCP unit cell
$V_\mathrm{hcp}=\left(c/a\right)_\mathrm{opt}a_\mathrm{hcp}^3\sin(2\pi/3)$ the optimal lattice
constant $a_\mathrm{hcp}$ was determined for each value of the niobium
concentration in the alloy. The obtained results for the lattice constant $a_\mathrm{hcp}$
and the structural relation $c/a$ as functions on the niobium concentration $x$
in the HCP Zr--$x$Nb alloy are shown in figures~\ref{fig2}~(a) and \ref{fig2}~(b),
respectively.

In figures~\ref{fig2}~(a) and \ref{fig2}~(b), the obtained theoretical results are shown
by circles. The values for the pure HCP zirconium are shown  by symbol $\times$.
First, let us consider the dependence of the lattice constant $a_\mathrm{hcp}$ versus
niobium concentration shown in figure~\ref{fig2}~(a). It is seen, that an increase in
the niobium concentration $x$ in the HCP alloy Zr--$x$Nb leads to a decrease in
the lattice constant $a_\mathrm{hcp}$, like in the case of the BCC crystals. However,
in contrast to the previous case, the descending dependence
$a_\mathrm{hcp}(\textrm{at}.\%\textrm{Nb})$ differs from the linear one for the shown interval of the
niobium concentration values. At low values of niobium concentrations
($\textrm{at}.\%\textrm{Nb}<20\%$), the calculated data for the lattice constant $a_\mathrm{hcp}$ lie on the
linear dependence, but for large at.$\%$Nb, one has a deviation from the linear
law. Thereby, the empiric Vegard's rule for the  HCP alloys can be applied
in the case of low values of niobium concentration $x$ in  Zr--$x$Nb alloys. In
the insertion in figure~\ref{fig2}~(a), we have shown the obtained theoretical results for
$a_\mathrm{hcp}$ at low at.$\%$Nb together with the experimental data presented by
triangles and squares from~\cite{fig1Ref2} and~\cite{fig2Ref1}, respectively.
One can see a good quantitative correspondence with experimental results for
small values of niobium concentration.

Next, let us consider the dependence of the structural relation $c/a$ to be a function
on the niobium concentration $x$ for the HCP Zr--$x$Nb alloys shown in
figure~\ref{fig2}~(b). Here, the markings are made in the same way as in figure~\ref{fig2}~(a)
for the lattice constant $a_\mathrm{hcp}$. It is seen that an increase in the niobium
concentration in the alloy leads to a growth of the structural relation $c/a$. This
means that the unit cell becomes more prolongate in $z$-direction. As for the
lattice constant $a_\mathrm{hcp}$, the dependence of the structural relation $c/a$
versus niobium concentration differs from the linear law; it can be applied
only for small values of at.$\%$Nb. Comparison of the results obtained for HCP
Zr--Nb alloy with experimental data from reference~\cite{fig1Ref2,fig2Ref1} for
structural relation $c/a$ at at.$\%\textrm{Nb}<10\%$ shows small deviation, which is
less than $1\%$ [see insertion in figure~\ref{fig2}~(b)].

Experimental investigations at $T\sim700$~K indicate that HCP phase in Zr--Nb
alloys is realized only if at.$\%\textrm{Nb}<6.5\%$~\cite{fig1Ref2,fig2Ref1}. Next,
analyzing the values of the total energy, the maximal niobium concentration value in
 Zr--Nb alloy is defined, when the alloy is still characterized by HCP lattice at low (zero)
temperature.

\section{Total energy and electronic density of states}

We have previously  calculated the optimal values for lattice constants for
Zr--$x$Nb alloys with BCC and HCP lattices at different values of niobium concentration
$x$. Since at normal conditions pure zirconium is characterized by the HCP
lattice, while pure niobium has BCC lattice, one can expect that at fixed
temperature there is a critical value for niobium concentration $x_\mathrm{c}$
corresponding to the HCP~$\rightarrow$~BCC structural transformation. This means
that for  Zr--$x$Nb alloy at $x<x_\mathrm{c}$, the HCP structure is  energetically more favorable, whereas at $x>x_\mathrm{c}$, the  BCC structure is realized. Experimental
investigations of zirconium-niobium alloys at high temperatures
 ($T\sim700$~K) show that in such conditions
the critical niobium concentration lies in the interval from 7 to $10\%$~\cite{fig1Ref1,fig1Ref2,fig2Ref1}.

To define the critical value of the niobium concentration $x_\mathrm{c}$ at low (zero)
temperature we analyze the total energies of all the crystals studied  with
both HCP and BCC lattices and determine the energetically most favorable
structure (HCP or BCC) for each model. To this end, we proceed in the following
manner. For the fixed niobium concentration $x$, using the obtained optimal lattice
constants $a_\mathrm{bcc}$, $a_\mathrm{hcp}$ and $c_\mathrm{hcp}$,  we perform a full cycle of
\emph{ab-initio} calculations to define the total energy for the unit cell.
In our calculations we study the unit cell containing 32 atoms: there is 1 atom of niobium and 31 atoms of zirconium for Zr--$3.125\%$Nb and one has 31 atoms of niobium and 1 atom of zirconium for Zr--$96.875\%$Nb.
 Next, we define the difference between the total energies for the
unit cell of 32 atoms of HCP and BCC lattice, as:
\[\Delta
E_\mathrm{tot}(\textrm{at}.\%\textrm{Nb})=E_\mathrm{tot}^\mathrm{uc}(\textrm{at}.\%\textrm{Nb})\big|_\mathrm{hcp}
-E_\mathrm{tot}^\mathrm{uc}(\textrm{at}.\%\textrm{Nb})\big|_\mathrm{bcc}\,.
\]
%
\begin{wrapfigure}{i}{0.5\textwidth}
\centering
 \includegraphics[width=0.45\textwidth]{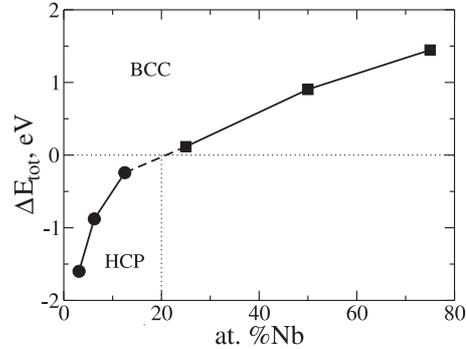}
\caption{The difference in the total energy of HCP and BCC $\Delta
E_\mathrm{tot}(\textrm{at}.\%\textrm{Nb})$ structures  in eV for the unit cell of 32 atoms as a function of the
niobium concentration in percentage for alloys Zr--Nb.} \label{fig3}
\end{wrapfigure}
Hence, the value $\Delta E_\mathrm{tot}(\textrm{at}.\%\textrm{Nb})$ defines the type of energetically
most favorable lattice for the Zr--$x$Nb alloy at fixed value of niobium
concentration $x$ as follows: (i) if
$E_\mathrm{tot}^\mathrm{uc}(\textrm{at}.\%\textrm{Nb})\big|_\mathrm{hcp} < E_\mathrm{tot}^\mathrm{uc}(\textrm{at}.\%\textrm{Nb})\big|_\mathrm{bcc}$,
and as a result $\Delta E_\mathrm{tot}(\textrm{at}.\%\textrm{Nb}) < 0$, then  Zr--$x$Nb alloy has HCP
lattice; (ii) at
$E_\mathrm{tot}^\mathrm{uc}(\textrm{at}.\%\textrm{Nb})\big|_\mathrm{hcp} > E_\mathrm{tot}^\mathrm{uc}(\textrm{at}.\%\textrm{Nb})\big|_\mathrm{bcc}$,
which yields  $\Delta E_\mathrm{tot}(\textrm{at}.\%\textrm{Nb})>0$, and the BCC lattice is realized. Therefore,
the critical value of the niobium concentration $x_\mathrm{c}$ [or $(\textrm{at}.\%\textrm{Nb})_\mathrm{c}$] that
defines the structural transformation HCP~$\rightarrow$~BCC at fixed
temperature (low/zero temperature in the studied case) can be determined from the
condition of energetic equivalence of two lattices, i.e., $\Delta
E_\mathrm{tot}(\textrm{at}.\%\textrm{Nb})=0$. The obtained results for the difference of the total energies
of the unit cell of 32 atoms versus niobium concentration in Zr--Nb alloy are
shown in figure~\ref{fig3}.

\begin{figure}[!b]
\centering
\includegraphics[width=0.45\textwidth]{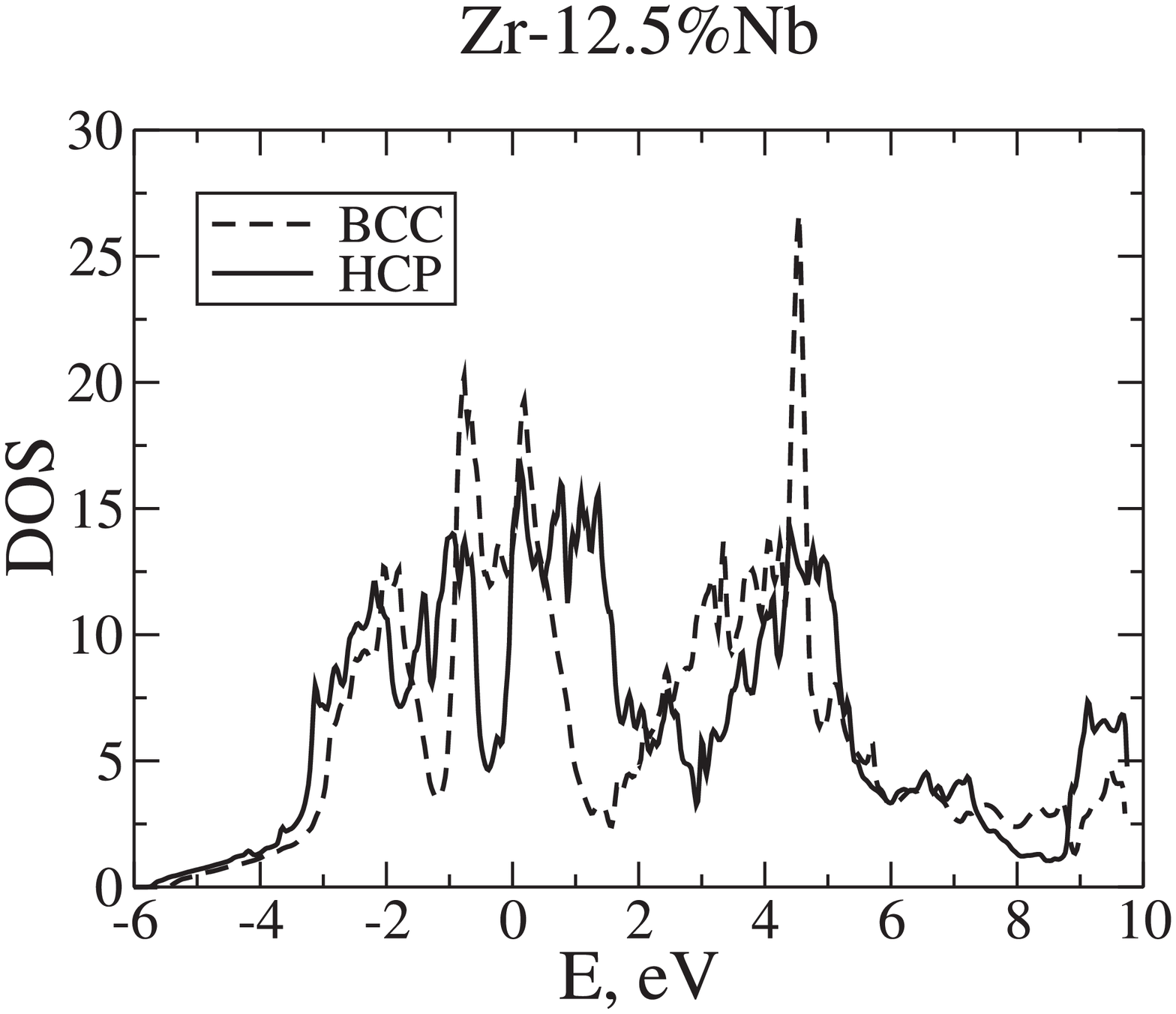}\hspace{0.05\textwidth}
\includegraphics[width=0.45\textwidth]{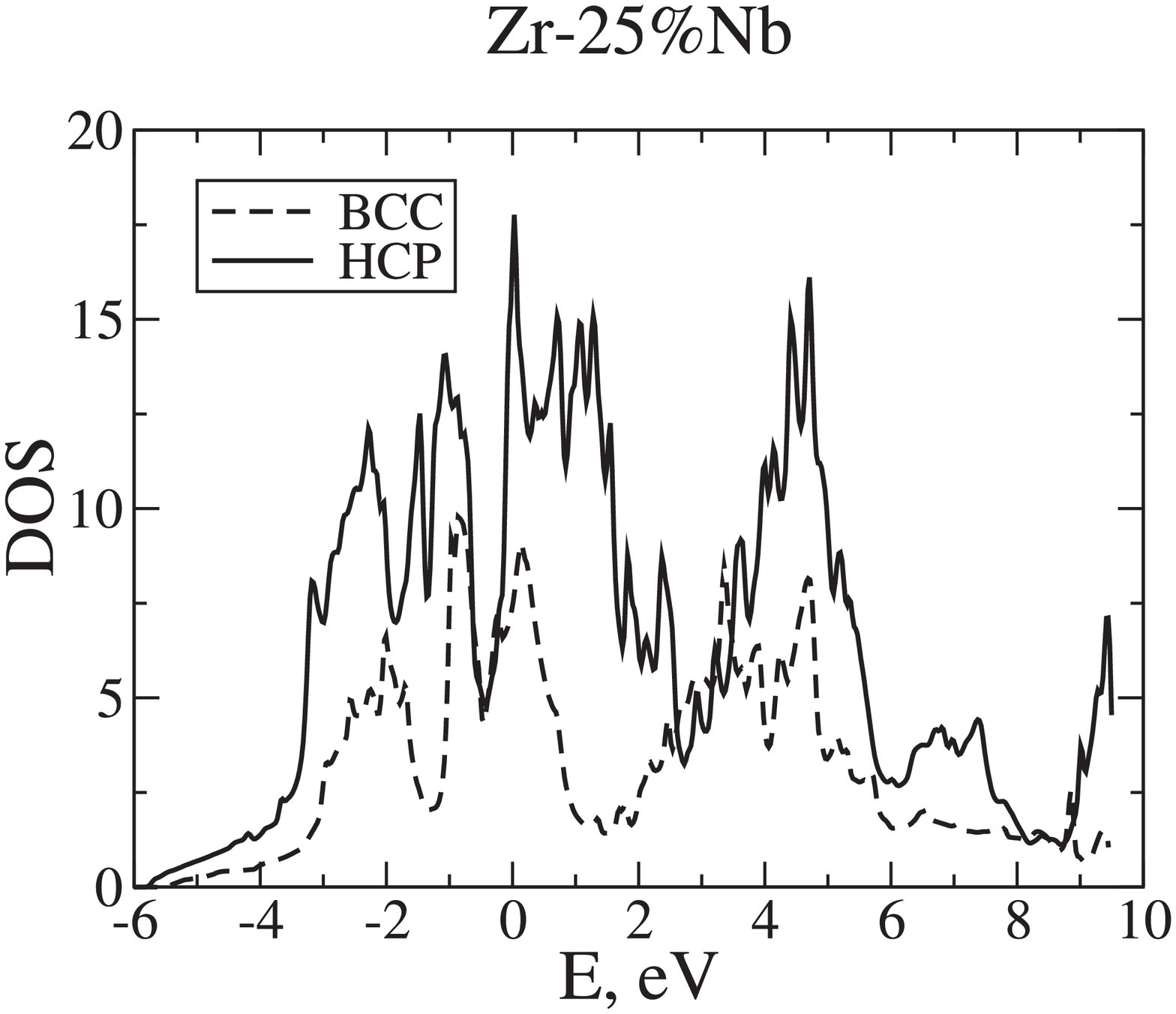}\\
\hspace{0.05\textwidth} (a) \hspace{0.5\textwidth}(b)
\caption{Dependencies for electronic densities of
states (DOS) versus the energy measured from the Fermi energy for: (a)
Zr--$12.5\%$Nb alloy; (b) Zr--$25\%$Nb alloy in the case of HCP lattice (solid curves)
and BCC lattice (dashed curves).} \label{fig4}
\end{figure}

It is seen that $\Delta E_\mathrm{tot}$ changes the sign from minus to plus in the
vicinity of the critical value of niobium concentration
$(\textrm{at}.\%\textrm{Nb})_\mathrm{c}\simeq20\%$. Therefore, at low (zero) temperature, the Zr--Nb alloy
with the niobium concentration less than $20\%$ is characterize by the HCP
lattice ($\Delta E_\mathrm{tot}<0$); the values for the difference $\Delta E_\mathrm{tot}$ are
shown by circles. In the case of large niobium concentration, i.e., over $20\%$,
the zirconium-niobium alloy has BCC lattice (squares in the figure~\ref{fig3}).
The obtained result provides an explanation as to the deviation from the linear
dependence of the obtained results for the lattice constant $a_\mathrm{hcp}$ of the
HCP alloys versus niobium concentration for large at.$\%$Nb [see figure~\ref{fig2}~(a)].

As was pointed out earlier, experimental investigations for the
zirconium-niobium alloys at high temperatures ($\sim700$~K) show that the BCC
phase is realized if $\textrm{at}.\%\textrm{Nb}\gtrsim10\%$, whereas the HCP phase can be
observed if $\textrm{at}.\%\textrm{Nb}\lesssim6.5\%$~\cite{fig1Ref1,fig1Ref2,fig2Ref1}. Thus,
the theoretically obtained result from \emph{ab-initio} calculations for the
critical value of niobium concentration in the Zr--Nb alloy
$(\textrm{at}.\%\textrm{Nb})_\mathrm{c}\simeq20\%$, which corresponds to the structural
HCP~$\rightarrow$~BCC transformation is about twice the experimental one. An increase of
the critical value of the niobium concentration with a decrease in the
temperature seems to be natural. With the temperature increase, one gets
thermal fluctuations becoming large in the vicinity of the lattice knots. It
leads to the destruction of the more complex HCP structure at a smaller niobium
concentration. Hence, one can expect a descending dependence of the critical
niobium concentration $(\textrm{at}.\%\textrm{Nb})_\mathrm{c}$ in Zr--Nb alloy versus temperature. The
character of the change in the critical value $(\textrm{at}.\%\textrm{Nb})_\mathrm{c}$ for the
HCP~$\rightarrow$~BCC structural transformation with an increase in the
temperature can be found using the molecular dynamics.

As far as the obtained critical niobium concentration at low (zero) temperature is
$(\textrm{at}.\%\textrm{Nb})_\mathrm{c}\simeq20\%$, next we perform calculations of the electronic density
of states (DOS) for two structures: Zr--$12.5\%$Nb and Zr--$25\%$Nb, which are
characterize by different energetically more favorable lattices, HCP and BCC,
respectively. The obtained results are shown in figure~\ref{fig4}.
Here, solid curves correspond to alloys with HCP lattices, whereas dashed ones
relate to alloys with BCC lattices. It is seen that for the
Zr--$12.5\%$Nb alloy the amplitude of the main peaks in the electronic density of
states is larger for the BCC lattice [see figure~\ref{fig4}~(a)], whereas for the
 Zr--$25\%$Nb structure, the situation is quite different: larger DOS corresponds to
the alloy with HCP lattice [see figure~\ref{fig4}~(b)]. These results prove that at
low (zero) temperature, Zr--$12.5\%$Nb alloy is characterized by the HCP lattice,
whereas alloy Zr--$25\%$Nb has BCC lattice.

\section{Conclusions}

We have studied structural properties of the zirconium-niobium alloys with
different niobium concentrations within the framework of \emph{ab-initio}
calculations. Niobium concentration varied from $3.125\%$ to $96.875\%$ for BCC
and HCP lattices.

Using the optimization procedure, optimal values for the lattice constants
were obtained. We have found that the introduction of zirconium atoms, as
substitutional ones, into the BCC crystal of niobium leads to an increase in the
lattice constant. The obtained dependence of a lattice constant versus zirconium
concentration in the BCC niobium alloy is in good agreement with empiric
Vegard's law and is quantitatively consistent with the known experimental data.
The introduction of substitutional niobium atoms into HCP zirconium crystal causes a decrease in the lattice constant and, as a result, to an increase in the
structural relation $c/a$. We have determined that at low (zero) temperature, the
critical value of the niobium concentration in Zr--Nb alloy, which defines the
structural HCP~$\rightarrow$~BCC transformation, is $(\textrm{at}.\%\textrm{Nb})_\mathrm{c}\simeq20\%$.

\ukrainianpart

\title{Структурні  властивості стопів Zr--Nb: моделювання \\ з перших принципів}
\author{В.О.Харченко\refaddr{label1,label2}, Д.О.Харченко\refaddr{label1}}
\addresses{
\addr{label1} Інститут прикладної фізики НАН України,  вул. Петропавлівська, 58, 40030 Суми, Україна
\addr{label2} Інститут фізики Аугсбургського університету,  вул. Університетська, 1, 86135 Аугсбург, Німеччина
}

\makeukrtitle

\begin{abstract}
\tolerance=3000%
Проведено першопринципні дослідження структурних властивостей стопів Zr--Nb із
довільними значеннями концентрації ніобію при нульовій температурі. Розглянуто
енергетичні властивості відповідних ОЦК та ГПУ структур. Отримано оптимальні
значення параметрів ґратки структур. Встановлено критичне значення концентрації
ніобію у цирконій-ніобієвих стопах, що відповідає структурному перетворенню
ГПУ~$\rightarrow$~ОЦК. Наведено густини станів для двох структур із
концентрацією ніобія $12.5\%$ та $25\%$, що характеризуються ГПУ та ОЦК
структурами, відповідно.
\keywords першопринципні розрахунки, ГПУ та ОЦК структури, стопи цирконій-ніобій

\end{abstract}

\end{document}